\documentclass[preprint2]{aastex}

\slugcomment{To submitted to ApJ Letters}

\shorttitle{NGC 6822}
\shortauthors{Demers et al.}

\begin{document}

\title{A Local Group Polar Ring Galaxy: NGC 6822\altaffilmark{1,2,3}}


\author{Serge Demers} 
\affil{Department of Physics, Universit\'e de Montr\'eal, Montr\'eal,
H3C 3J7, Canada}
\email{demers@astro.umontreal.ca}

\author{Paolo Battinelli}
\affil{INAF, Osservatorio Astronomico di Roma, I-00136, Roma, Italy}
\email{battinel@oarhp1.rm.astro.it}
\and 
\author{William E. Kunkel}
\affil{The Observatories of the Carnegie Institution of Washington,
        La Serena, Chile}
\email{kunkel@jeito.lco.cl}

\altaffiltext{1}{Based on observations obtained at the Italian Telescopio
Nazionale Galileo operated on the Island of La Palma by the Fundaci\'on
Galileo Galilei of the INAF at the Spanish Observatory de Roque de los
Muchachos of the Institute de Astrofisica de Canarias.}
\altaffiltext{2}{Based on observations obtained with MegaPrime/MegaCam, 
a joint project of CFHT and CEA/DAPNIA, at the Canada-France-Hawaii 
Telescope (CFHT) which is operated by the National Research
Council (NRC) of Canada, the Institut National des Sciences de l'Univers
of the Centre National de la Recherche Scientifique (CNRS) of France,
and the University of Hawaii.}
\altaffiltext{3}{Based on observations acquired at the du Pont
Telescope, Observatories of the Carnegie Institution of Washington.}


\begin{abstract}
Star counts, obtained from a 2 $\times$ 2 degree area centered on NGC 6822
have revealed an optical image of this galaxy composed of two components:
in addition to the well-known HI disk with its young stellar component, there 
is a spheroidal stellar structure as extensive as its HI disk but with its 
major axis at roughly right angles to it which we traced to at least 36
arcmin.  Radial velocities of over 100 
intermediate-age carbon stars found within this structure display kinematics 
contrasting strongly with those of the HI disk. These C stars belong to the 
spheroid. Although devoid of gas,
the spheroid rotation is consistent with the I-band Tully-Fisher relation.
 The orientation of the 
rotation axis which minimizes the stellar velocity dispersion coincides with 
the minor axis of the stellar population ellipsoid, lying very nearly in the 
plane of the HI disk.  We conclude: that the HI disk is a polar ring and the
spheroidal component an erstwhile disk, a  
fossil remainder of a past close encounter episode.

\end{abstract}


\section{Introduction}

NGC 6822 is a typical Magellanic dwarf Irregular  galaxy located
at 500 kpc from the Sun, and after the Magellanic Clouds, is our closest 
neighbor more luminous than M$_V$ = --16. 
The optical appearance of this galaxy is dominated by a bar, mapped
by 
\citet{hod77}, about 8$'$ long and with a position angle (PA)of 10$^\circ$
while its HI content forms a huge disk at a PA $\sim$ 130$^\circ$ \citep{wel03}.
Dwarf Irregular galaxies (dIrrs) and dwarf spheroidal/elliptical 
galaxies (dSph/dEs)  
differ primarily by their baryonic content, with the virtual 
absence of a stellar population younger than a crossing time
and no significant HI disk in dSph/dEs; the known exceptions, such as NGC 205, 
bear marks of a tidally disturbed past.  
dIrrs, on the other hand, 
are marked by a photometrically dominant population of young stars
unevenly distributed over the inner portions of a pronounced HI disk that 
often extends (at 10$^{19}$ cm $^{-2}$) well beyond their Holmberg radii.
In terms of their surrounding environment the dSph/dEs are found concentrated 
toward regions of high galaxy density in clusters where mean times to 
harassing encounters are significantly less than a Hubble time.  The dIrrs 
are found preferentially in isolated environments where the mean time for 
a tidally harassing encounter is long, so that for their intrinsic observed 
velocities the likelihood of encountering a ``neighbor'' of comparable or 
greater mass in less than a Hubble time is esteemed to be minute.  

A currently fashionable scenario envisions these two classes as sharing a 
common origin, with transitions of dIrrs and small spirals into 
dSph/dEs types arising from ram
pressure stripping \citep{fab83} or galaxy harassment \citep{moo98}
by more massive neighbors.
With the long relaxation time of dwarf galaxies, the former mechanism
does not significantly modify the disk angular momenta while the latter
may do so when a sufficient number of harassments accumulates.

It is worth contrasting this view with an alternate relying on
numerical simulations initiated by Alar and Juri Toomre 
\citep{too72}. These demonstrate that tidally generated
tails of interacting galaxies may leave debris mixtures of some variety.
The internal dynamical signatures of this debris, such as spin, 
velocity dispersion and orbital 
angular momentum would almost certainly be distinct 
from those of dwarf systems of primordial
origin.  Observations suggesting the current formation of such stellar 
debris during a tidally disruptive encounter are well known 
\citep{sch83,bar92}.

Observational properties of a dwarf galaxy are not unambiguously
indicative of which formation mechanism is relevant. 
Consequently it becomes risky to 
propose cosmological inferences from dynamical traits observed in dwarf
galaxies
whose dynamical histories are no longer uniquely determined.  The most
irritating aspect of this conundrum is the difficulty in accounting for
the presence or absence of system spin.

Recent surveys, over large angular areas, of Local Group galaxies have 
revealed that these galaxies, being spiral or dIrrs
are much bigger than previously thought. M31's halo and disk have recently
expanded \citep{gah05, iba05}, the disk of NGC 300 
has been detected up to
10 scale lengths \citep{bla05}. Our Local Group carbon
survey reveals the existence of two distinct scenarios: 1)
a stationary  environment, i.e., WLM, NGC 3109, NGC 185
\citep{bat04a,dem03,bat04b}
where all the C stars lie within 
a few scale lengths from the center; 2) scenarios where we see
dynamical violence in their past history. The 
primary examples of the latter are the Magellanic Clouds \citep{irw91},
NGC 6822 
\citep{let02} and IC10 \citep{dem04} 
with carbon stars found beyond seven scale
lengths. The LMC is quite extended with fragments of the disk seen 
by \citet{gal04} 7$^\circ$ to the north.

For the last twelve years we have identified C stars in several
Local Group galaxies (see \citet{bat05} for a summary)
for eventual
use as dynamical test particles in the outer parts of galaxies
\citep{kun97a,kun97b,kun00},
 to facilitate
tracking of angular momentum. 

\section{Observations}
The photometric data set, discussed in this letter, corresponds to r$'$ and
i$'$ images taken with MegaPrime/Megacam in Queue
Observing mode in May and June 2004 on the Canada-France-Hawaii Telescope. 
The wide field imager Megacam consists of 36 2048 $\times$ 4612 pixel
CCDs, covering nearly a full 1$^\circ \times 1^\circ$ field. 
It offers a resolution of 0.187 arcsecond per pixel.
Four slightly overlapping fields were observed, the galaxy located
at the common corners, to cover essentially a 2$^\circ \times 2^\circ$ area
centered on NGC 6822.

The data distributed by the CFHT have been pre-reduced, corrected for
bias, flat fielded, etc. 
The photometric reductions were done by Terapix, the data reduction
center dedicated to the processing of extremely large data flow. The Terapix
team, located at the Institut d'Astrophysique de Paris, matches and stacks all
images taken with the same filter and, using SExtractor \citep{ber96},
provided magnitude calibrated
catalogues of objects in each of the combined images.

The spectroscopic data consist of two sets of observations. Spectra
of carbon stars (identified by  \citet{let02})
were obtained with two telescopes.
Fifty stars were observed with the WFCCD spectrograph, in 
its \'echellette mode,
 at the du Pont 2.5 telescope of Las Campanas Observatory in August 2002. 
The FWHM resolution is 1.7\AA\  over the spectral region  7850 to 8760 \AA.
Sixty stars were observed with DOLORES multi-object spectrograph 
(FWHM = 3.1 \AA) attached to the
Telescopio Nazionale Galileo located on Cerro de Los Muchachos, La Palma. 

\section{Isodensity map}
The surface density of stars in the direction of NGC 6822 is  high
because of its relatively low Galactic latitude (b = --18.4$^\circ$). 
In order to increase the
contrast between its low density periphery and the foreground field,
we selected only stars with (r$'$ -- i$'$)$_0$ colors and magnitudes
corresponding to the NGC 6822 red giant branch (RGB) stars.  The reddening
of the entire field has previously been mapped to deredden each star
\citep{dem05}.  This giant star selection yields
some 150,000 stars, most of them members of NGC 6822.
The whole field is then covered by a 50 $\times$ 50 pixel wide grid
and stars are counted  over a circular 500 pixel sampling area, centered
on each intersection of the grid. This is done to  smooth out major
irregularities (mainly due to bright foreground stars that locally prevent
the detection of fainter members of NGC6822). This density map is
transformed into a density image that is analyzed with 
IRAF/STSDAS/analysis/isophote/ellipse
to fit isodensity ellipses, determine their position angles and
ellipticities. This technique has previously been employed by us to
map the structure of IC 10 \citep{dem04}.

Figure 1 presents the isodensity contours of the RGB stars. Contours
corresponding to 3,10,20,30 sigmas above the average count level are shown.
The inner ellipse fits the 3$\sigma$ contours while the larger one
corresponds to 1.2$\sigma$ and is the outermost one identified by the
IRAF ELLIPSE task, it has a semi-major axis of 36$'$.
One can see that the major axis of the ellipse is nearly orthogonal
to the HI disk, represented by the dashed line.
This spheroid density profile is well fitted by a 
 two-exponential law with scale 
lengths of 3.8$'$ and 10$'$, which are interpreted as follow.  
Beyond the Freeman radius
of roughly 5 scale lengths (of 3.8$'$) we perceive a change in slope, 
a flattening,
that in very roughly corresponds to the radius where tidal deformations
from interactions might be expected.  N-particle simulations should
eventually clarify this interpretation.
The two ellipses in Fig. 1
are characteristic of the whole family of ellipses of various
major axes than can be traced. Indeed, from 10$'$ to 35$'$ the
PA of the major axis varies from 80$^\circ$ to 65$^\circ$ while
the ellipticity range from 0.24 to 0.38. More details can be found
in our forthcoming paper \citep{dem05}.
 This figure shows that the bulk of the spheroidal population stars in NGC 6822,
surrounding its bright central bar is comparable in size to the HI disk,
mapped by \citet{deb00},  but oriented quite differently.

\section{Kinematics}

Radial velocities of 110 carbon stars observed within 15$'$ of the HI major
axis from an earlier survey \citep{let02} are described here.  The
spectra covered the spectral domain from 7500 to roughly 9000 \AA, relying on
roughly 50 night sky lines for wavelength calibration, and four template
carbon stars from \citet{tot98}; velocity variability affect
these N-type carbons, limiting the system precision to $\pm$15 km s$^{-1}$.
Their
mean distribution, referred to TI 0357+0908, lies between +10 and 
--70 km s$^{-1}$. Figure 2 shows the radial velocities plotted as a function
for the spheroid major axis.
Even so, carbon stars as kinematic ``test particles'' airport the advantage
of freedom from contamination the Galactic foreground might otherwise
impose.  Telluric absorption features (from primarily H$_2$O and O$_2$) were
used to compensate for possibly uneven illumination of the slit masks that
might otherwise introduce an arbitrary velocity shift.

         The spatial distribution of these carbon stars fits the spheroid
better than the HI.  Quite apart from this spatial distribution, however,
it was found that the coordinate system x', y' that yielded the minimal
residuals in rotation velocities is directed toward PA of between 63 and
67 degrees, and so places the rotation axis of the system of carbon stars
at very nearly right angles to that of the HI disk and close to the minor
axis of the spheroid. Figure 3 shows the variation of the dispersion
residuals with trial orientations of the x', y' frame.  Our conclusion is
that the carbon stars (1) show no preference for the HI disk and (2)
demonstrate that they form part of a stellar population rotating at nearly
right angles to the HI disk, leading to the suggestion that the HI disk is
a structure more reminiscent of a polar ring.

\section{DISCUSSION}

The close similarity between the morphology and kinematics of NGC 6822 and
classical polar ring galaxies (PRG's) suggests analogous formation
scenarios, which we explore briefly (with our thanks to the referee).
Foremost among these similarities are the nearly perpendicular
orientations of two systems of angular momenta.  Also, investigators of
PRG's have commented that almost all such systems show indications of
recent formation, possibly within the last two Gyr \citep{whi90,iod03}.
Last, and from our perspective the most significant,
is the severe dissimilarity between the population types of these two
components:  one being virtually gas-free, while the other shows little
evidence for the existence of an older component (represented by an RGB).
More explicitly, the absence of carbon stars in the HI disk on the one
hand, and simultaneously the absence of gas and a young stellar population
in the spheroid on the other carries important connotations for time
scales.

 The presence of carbon stars in a dominated by RGB stars 
spheroidal component of NGC 6822 bears similarities with other
spheroids of the Local Group \citep{bat05} even though
major differences must be noted.  No other spheroid of the Local
Group shows comparable spin;  for NGC 205 (M$_V$ = --16.3) the spin found by
\citet{geh05} is one seventh that detected here, and a comparison
with dwarf ellipticals of the Virgo Group by \citet{geh03} places the
spheroid of NGC6822 far above the upper limit of their sample (their Fig.
4). Comparing the luminosity function of the spheroid's RGB with that of
Fornax \citep{dem94} yields an M$_V$ = --14.6, or M$_I$ = --15.8
then locates the spin of NGC6822's spheroid
within one standard deviation of the infrared Tully-Fisher relation as
given by \citet{gio97}.  This suggests interpretation that, instead of a spheroid,
 we are seeing a
dwarf disk that has lost its original gaseous and young population
component.

An inspection of the 
 distribution of the young population associated with the HI disk may help
to put constraints to possible scenarios. These stars fill
a narrow elongated ellipse close to the outer HI periphery \citep{bat03}.  
The outer
envelope of this distribution is the same, and comparably populated, for
stars of $\sim$ 500 Myr age (as determined from isochrone fittings)
as is found
for extremely young stars of $\sim$ 50 Myr, suggesting that the last traces
of transient tidal phenomena had extinguished completely at some prior epoch,
leading to the conclusion that whatever tidal event produced the polar ring
phenomenon must be significantly older. 

In their descriptions of the classical PRG's \citet{whi90} comment
on the population dissimilarity between the two components.  We note also
an observation from \citet{too72} that strikes us as crucially
relevant. To start a merger or accretion process that initiates a hiatus or
break of duration longer than a crossing time of the spheroid or the polar 
ring without going to completion requires incredibly finely tuned 
dynamical starting conditions of low orbital energies.  Such low 
orbital energy differences between two neighboring galaxies in a cluster 
such as where S0 galaxies are observed is an exceedingly unlikely 
occurrence.
More problematic, if what is seen are incomplete mergers,
why are such detained mergers observed only when one member (generally the
more massive) is an almost gas-free system, not far in the Hubble sequence
from the S0 types, and the other always gas rich?  If the orbital
energetics of such finely tuned merging encounters are the determining
factor for the formation of PRG's, then all Hubble types should be
represented, and this is not seen.

From our work of carbon stars as dynamical test particles in interacting
disruptive encounters between the Magellanic Clouds, another formation
scenario for PRG's,  not treated by \citet{bou03} in their
exploration of likely formation processes, strikes us as capable of
generating ``gentle'' merger conditions thought significant by the Toomre's
and simultaneously to account for the population dissimilarity between the
two components.  In this concept a disk galaxy striking the outer envelope
of a more massive perturber experiences a purely gravitational though
relatively mild impulse to its stellar component, but additionally another
incremental impulse to its gaseous component.

In the case of the
SMC traversing the LMC disk decreased the orbital angular momentum of the
gaseous component more so that the purely gravitational tidal impulse, so
as to lag behind the stellar component by almost 2 kpc some 300 Myr later
\citep{kun00}; other manifestations separating the gas component
trajectories from those of the stellar are apparent in much of the region
between the Magellanic Clouds.  We recall the case of UGC7636, a dwarf
galaxy which, after interacting with the elliptical NGC 4472 left its
entire gas component several system diameters behind in orbit \citep{san87}
a more aggravated interaction of similar sort.  Without
asking with which of the two separated components of UGC7636 a DM halo
might remain, we note that as an instance scaled somewhere in between the
SMC episode and that of UGC7636, the PRG's syndrome would appear best
represented by an incomplete tidal disruption in which the PRG hiatus is
now more easily placed into context phenomenologically, accounting
entirely for the dissimilarity between the population contents of the two
components.

We conclude noting that whichever scenarios for a PRG formation one prefers, 
encounters with massive neighbors are always needed. This is quite a puzzling
circumstance since NGC 6822 is considered to be a typical isolated dwarf galaxy.
Further investigations are needed to solve the riddle of the missing culprit. 
\acknowledgments
This research is funded in parts (S. D.) by the Natural
Sciences and Engineering Research Council
of Canada.

{\it Facilities:} \facility{CFHT (Megacam)}, \facility{Du Pont (WFCCD)}, 
\facility{TNG (DOLORES)}

\clearpage

\begin{figure}
\epsscale{0.80}
\plotone{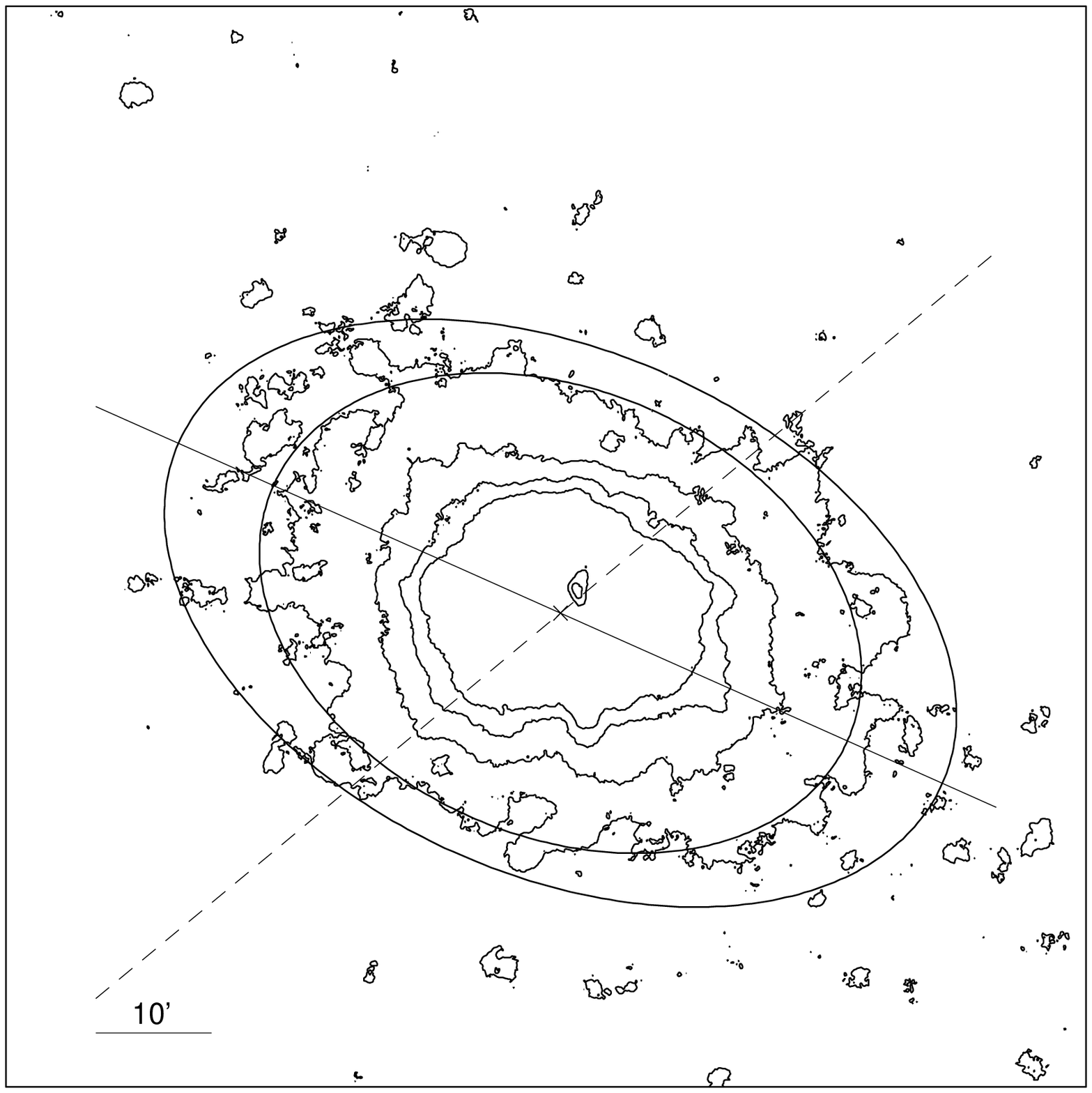}
\caption{Isodensity contours of the RGB stars, centered on NGC 6822.
The solid line represents the major axis of the outermost ellipse while the
dashed line shows the orientation of the HI disk. See text for further 
details. North on top and East to the left.
\label{fig1}}
\end{figure}

\clearpage

 \begin{figure}
\plotone{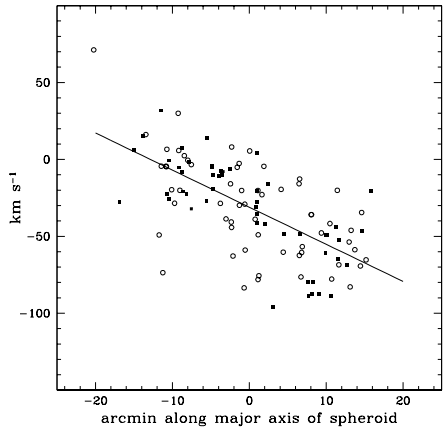}
\caption{Observed radial velocities of C stars, the axis selected
has a PA = 64.5$^\circ$. Solid dots are du Pont data while open circles
are from the TNG. \label{fig2}}
\end{figure}

\clearpage

\begin{figure}
\epsscale{0.80}
\plotone{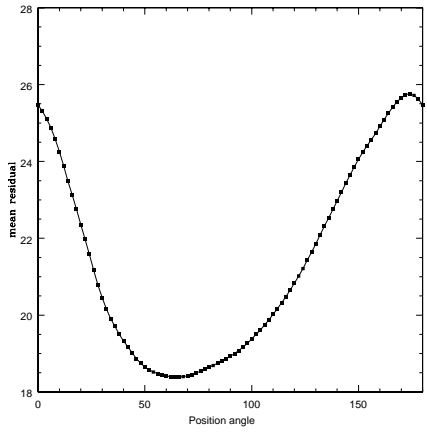}
\caption{The mean residuals of the radial velocities of C stars for
various position angles of the major axis.
\label{fig3}}
\end{figure}


\begin{thebibliography}{}

\bibitem[Barnes \& Hernquist(1992)]{bar92} Barnes, J. E. \& Hernquist, L.
1992, Nature, 360, 715

\bibitem[Battinelli et al.(2003)]{bat03} Battinelli, P. Demers, S, \&
Letarte, B. 2003, \aap, 405, 563

\bibitem[Battinelli \& Demers(2004a)]{bat04a} Battinelli, P. \& Demers
S. 2004a, \aap, 416, 111

\bibitem[Battinelli \& Demers(2004b)]{bat04b} Battinelli, P. \& Demers
S. 2004b, \aap, 417, 479

\bibitem[Battinelli \& Demers(2005)]{bat05} Battinelli, P. \& Demers
S. 2005, \aap, 434, 657

\bibitem[Bertin \& Arnouts(1996)]{ber96} Bertin, E., \& Arnouts, S. 1996,
\aap, 117, 393

\bibitem[Bland-Hawthorn et al.(2005)]{bla05} Bland-Hawthorn, J., Vlaji\'c,
M., Freeman, K. C., Draine, B. T. 2005, \apj, 629, 239

\bibitem[de Blok \& Walter(2000)]{deb00} de Blok, W. J. G. \& Walter, F. 2000
\apj, 537, L98

\bibitem[Bournaud \& Combes(2003)]{bou03} Bournaud, F. \& Combes, F. 2003,
\aap, 401, 817

\bibitem[Demers et al.(1994)]{dem94} Demers, S., Irwin, M. J.,
\& Kunkel, W. E. 1994 \aj, 108, 1648

\bibitem[Demers et al.(2003)]{dem03} Demers, S., 
Battinelli, P., \& Letarte, B. 2003, \aap, 410, 795

\bibitem[Demers et al.(2004)] {dem04} 
Demers, S., Battinelli, P., \& Letarte, B. 2004, \aap, 424, 125

\bibitem[Demers et al.(2005)] {dem05} Demers, S., Battinelli, P.,
\& Kunkel, W. E. 2005, in preparation

\bibitem[Faber \& Lin(1983)]{fab83} Faber, S. M., \& Lin, D. N. C. 1983, \apj,
266, 17

\bibitem[Guhathakurta et al.(2005)]{gah05} Guhathakurta, P., Ostheimer, 
J. C., Gilbert, K. M., Rich, M. R., Majewski, S. R., Kalirai, J., Reitzeil,
D. B., \& Patterson, R. J. 2005, preprint (astro-ph/0502366)

\bibitem[Gallart et al.(2004)] {gal04} Gallart, C., Stetson, P. B., 
Hardy, E., Pont, F., \& Zinn, R.  2004, \apj, 614, L109

\bibitem[Geha et al.(2003)]{geh03} Geha, M., 
Guhathakurta, P., van der Marel, R. P. 2003, \aj, 126, 1794

\bibitem[Geha et al.(2005)]{geh05} Geha, M., Guhathakurta, P., Rich, R. M.,
\& Cooper, M. C. 2005, \aj, in press (astro-ph 0509561)

\bibitem[Giovanelli et al.(1997)]{gio97} Giovanelli, R., Haynes, M. P. da Costa,
L. N., Freudling, W., Salzer, J. J., \& Wegner, G. 1997, \apj, 477, 1

\bibitem[Hodge(1977)] {hod77} Hodge, P. W. 1977, \apjs, 33, 69

\bibitem[Ibata et al.(2005)]{iba05} Ibata, R., Chapman, S., Ferguson,
A. M. N., Lewis, G, Irwin, M., \& Tanvir, N. 2005, prepint (astro-ph/0504164)

\bibitem[Iodice et al.(2003)]{iod03} Iodice, E., Arnaboldi, M., Bournaud, F.
et al. 2003, \apj, 585, 730

\bibitem[Irwin(1991)]{irw91} Irwin, M. J. 1991, in The Magellanic Clouds,
IAU Symp.
148. ed. R. Haynes and D. Milne, Kluwer, Dordrecht, 1991., p.453

\bibitem[Kunkel et al.(1997a)]{kun97a} Kunkel, W. E., Demers, S.,
\& Irwin, M. J. 1997a, \aaps, 122, 463

\bibitem[Kunkel et al.(1997b)]{kun97b} Kunkel, W. E., Demers, S., Irwin,
M. J., \& Albert, L. 1997b, \apj, 488, L129
 
\bibitem[Kunkel et al.(2000)]{kun00} Kunkel, W. E., Demers, S.,
Irwin, M. J. 2000, \aj, 119, 2789

\bibitem[Letarte et al.(2002)]{let02} Letarte, B., Demers, S., Battinelli,
 P., \& Kunkel, W. E. 2002, \aj, 123, 832

\bibitem[Moore et al.(1998)]{moo98} Moore, B., Lake, G., \& Katz, N.
1998, \apj, 495, 139

\bibitem[Sancisi et al.(1987)]{san87} Sancisi, R., Thonnard, N., \& Ekers,
R. D. 1987 \apj. 315, L42

\bibitem[Schweizer et al.(1983)]{sch83} Schweizer, F., 
Whitmore, B. C., \& Rubin, V. C. 1983, \aj, 88, 909

\bibitem[Toomre \& Toomre(1972)]{too72} Toomre, A. \& Toomre, J. 1972, \apj, 178,
623
\bibitem[Totten \& Irwin(1998)]{tot98} Totten, E. J. \& Irwin, M. J. 1998,
\mnras, 294, 1

\bibitem[Weldrake et al.(2003)]{wel03} Weldrake, D. T. F.,
de Blok, W. J. G., \& Walter, F. 2003, \mnras, 340, 12

\bibitem[Whitmore et al.(1990)]{whi90} Whitmore, B.C., Lucas, R.A.,
 McElroy, D.B., et al. 1990, \aj, 100, 1489

\end{thebibliography}
\end{document}